# Numerical Analysis Investigation of Acoustic Shadow Moiré Interference

Ibrahim Abdel-Motaleb[*] and Mahmoud Yaqoub
Department of Electrical Engineering, Northern Illinois University, DeKalb, IL 60115
[*]ibrahim@niu.edu

*Abstract-* **Acoustic shadow moiré has unique properties to be used for many potential applications in medical diagnostics, manufacturing, and material characterization. In this paper, numerical analysis, using Comsol, is used to investigate the principles of acoustic moiré interference phenomenon. The study confirmed that the expected fringe images of shadow interference can be created at Talbot distances. The study confirms the experimental results reported by this group [1] and proves without any doubt that acoustic moiré interference phenomenon exists.**

## I. INTRODUCTION

Optical Moiré effect is formed when light rays is applied on superimposed gratings. The commonly used grating patterns are lines, square meshes, dot patterns, concentric circles, radial lines, and spirals. In our study, only line gratings are used. Moire phenomenon is very sensitive to the slightest variation or distortion to the overlaid structure. This sensitivity gives a unique ability to detect defects, distortions, or asymmetries in the inspected object. Superimposing two gratings structures results in structures that are different from, but related to, the gratings. From this relationship, information about the imaged object can be obtained.

The Moiré effect occurs because of the interaction between the overlaid structures that interfere with each other to create inference patterns or moiré fringes [1,2]. An example of the shadow moiré interference is shown in Fig. 1. The grating is formed using straight lines, but the interference pattern is composed of bent and distorted lines, as shown in Fig. 1. The shape of the contours and their position are related to the out of plane elevation of the object surface, as it will be explained in details later. In other words, the surface topology can be obtained from the observed moiré image.

Imaging using moiré phenomenon can be done using several methods. One of them is called the out of plane method, where the reference grating is placed at a specific distance from the surface of the object. The shadow of the reference grating is laid on top of the object. Interference between the shadow and the actual grating creates interference pattern, as shown in Fig. 2. Another method is called the in-plane moiré method. In this method, two gratings are used. The first is a specimen grating, which is affixed to the surface of the object either by cementing or by printing. The second is a reference grating through which the specimen is viewed. Moiré fringes are caused by interference between the reference and the specimen grating. In this study the out of plane method is used.

Consider the out of place method, Fig. 2, the incident angle at point "Q" is $\alpha$ and the observation angle is $\beta$. However, for point "Q1" the incident light angle is $\alpha1$ and the observation angle is $\beta1$. Since $\alpha$ and $\beta$ are different from $\alpha1$ and $\beta1$, then the observed grating at "Q" and "Q1" would be different. This is why the observed grating may look like Fig. 1. From the values of the angles and the grating pitch, g, one can obtain the topography of the surface of the specimen, as explained below.

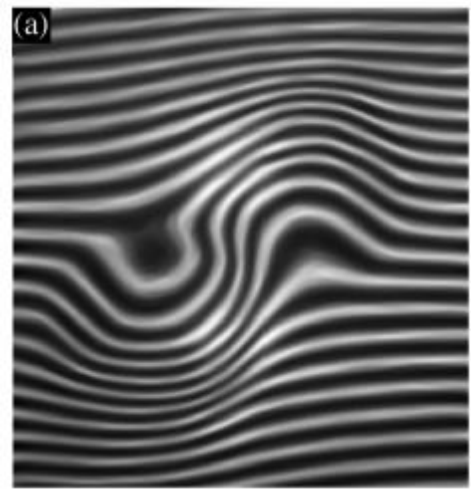

Fig. 1. Noiseless shadow moiré image. Image is used with permission from the IEEE [2].

Traditionally, optical waves are used in shadow moiré technique. Can acoustic waves be used too? Acoustic waves are longitudinal, compressional waves that can be periodic or pulsed. More generally, sound can refer to any type of mechanical wave motion, in a solid or fluid medium, which propagates via the action of elastic stresses, involving local compression and expansion of the medium [2]. The propagation of sound in fluids happens through pressure waves or compressible waves. In solids acoustic, waves propagate through small amplitude elastic oscillations of its shape. Adult humans can hear sounds at frequencies between about 20 Hz to 20 kHz. Ultrasonic, on the other hand, covers a huge range of frequencies above 20 kHz to the GHz range. Acoustic microscopy uses MHz to GHz range [3,4].

In this paper, we report on the numerical analysis investigation for acoustic moiré fringes using Comsol multiphysics program. The main goal is to prove that acoustic shadow moiré exists and its fringes are created as predicted experimentally.

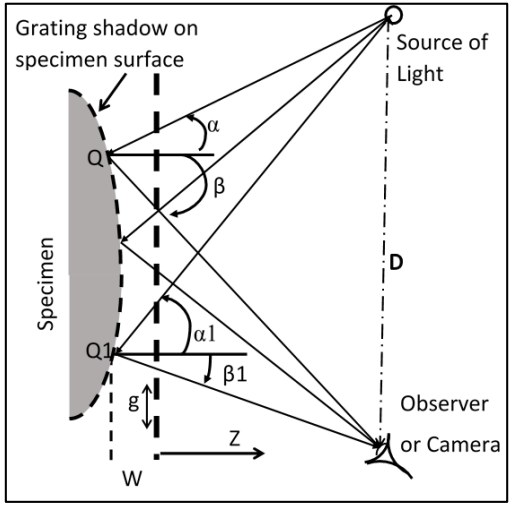

Fig. 2. Schematic of out of plane Shadow Moiré method; after [5].

## II. ANALYSIS OF SHADOW MOIRÉ

Shadow moiré method is used to measure out-of-plane displacements of a surface. This is achieved by obtaining the distance, z, between every point on the surface and a reference plane. This reference plane is the physical grating mask, as shown in Fig. 2. For point Q1 on the object surface, its distance z from the physical grating is equal to W. Now, if the distance between every point on the surface and the reference grating is obtained, the surface topology of the object can be obtained.

To obtain the equations that determine the distance, a detailed diagram for Point Q of Fig. 2, is shown in Fig. 3. From Fig. 3, the gap, z, between the surface (point Q) and the grating can be obtained from the relation [5]:

$$z = \frac{N \cdot g}{(\tan(\alpha) + \tan(\beta))} \quad (1)$$

where g is the pitch, N is the number of pitches between points A and B, $\alpha$ is the angle of the incident light, and $\beta$ is the angle of the observation line.

At point Q1, Fig. 2, the angles are $\alpha 1$ and $\beta 1$, hence z =W is

$$W = \frac{N \cdot g}{(\tan(\alpha 1) + \tan(\beta 1))} \quad (2)$$

In reality, Q is seen as point B by the observer or camera. To reduce this error, the gap between the surface and both the source and the camera must be very small. The vertical separation for one spacing between two moiré fringe is equal to C, where,

$$C = \frac{g}{(\tan(\alpha) + \tan(\beta))} \quad . \quad (3)$$

The intensity distribution of shadow moiré fringes formed by a monochromatic source light with wave length, $\lambda$, at a distance z from the grating (Fig. 2) can be expressed by [6,7]:

$$I_s = \frac{1}{4} + \sum_{n=1}^{\infty} \frac{2}{\pi^2 (2n-1)^2} \cos\left\{\frac{2\pi z (2n-1)^2}{D_T^\alpha}\right\} \cos\left\{\frac{2\pi z (2n-1)}{g/\tan\alpha}\right\} \quad (4)$$

Here z is the distance between the grating and the position at which the image is observed. $D_T^\alpha$ is the Talbot distance with incident light applied at an angle $\alpha$. $D_T^\alpha$ can be expressed as,

$$D_T^\alpha = \frac{2g^2 \cos^3 \alpha}{\lambda} \quad (5)$$

Talbot distance is the distance at which the shadow will be a repeat of the actual grating. Therefore, if it is found that the grating image is replicated at that distance, one can be certain that shadow moiré interference is created. Explanation of Talbot distance and images are reported in details in [1, 6,7]. From Eq. 4, the intensity of the fringes at any distance z can be obtained. If $z = 1 D_T^\alpha$, the created image will be a self-image of the grating. If $z = \frac{1}{2} D_T^\alpha$, the image will the complement of the grating. At $z = \frac{1}{4} D_T^\alpha$ and $z = \frac{3}{4} D_T^\alpha$, the images should have double the frequency of maxima. Hence, to prove that acoustic moiré interference takes place, one needs to capture the image of the acoustic wave at $1 D_T^\alpha$ or at $\frac{1}{2} D_T^\alpha$. Fig. 4 shows a general setup, where ultrasound waves are applied with an angle $\alpha$, and images are captured at ¼, ½, ¾, and 1 Talbot distances.

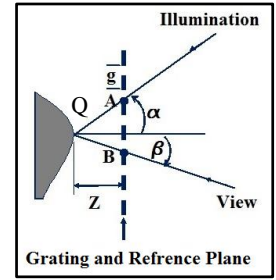

Fig. 3. Details at point Q of Fig. 2. Here Z is W in Fig. 2.

## III. SIMULATION RESULTS AND ANALYSIS

Comsol uses Finite Element method (FE) to solve the Partial Differential Equations (PDEs) with some boundary conditions to obtain the behavior of a physical system. The execution process starts by identifying the governing equations and deriving the necessary approximations. The system is then discretized into finite number of elements. A global matrix of equation is created from all elements and solved, within the accepted error. In each element, an approximation of the dependent variable is introduced. Then the integral form equation is evaluated for each element in order to assemble the solution as a global matrix equation. The said matrix equation is solved. Finally, from the solution of the matrix, values of interest can be calculated. To ensure accuracy, the mish size is chosen to be 0.2 of the wave length. The details of the simulation are explained in [8].

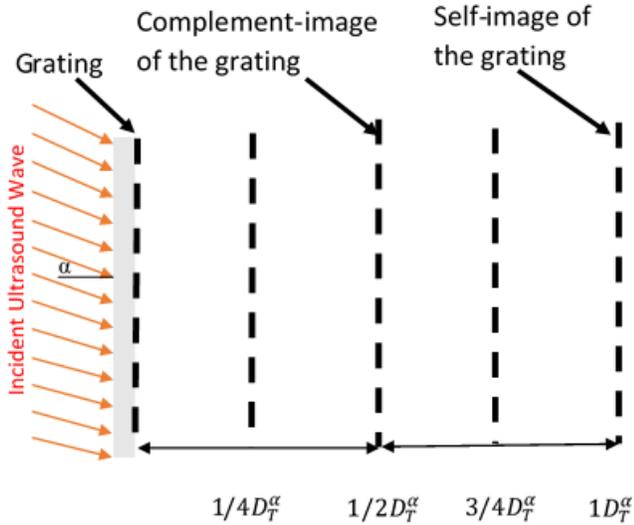

Fig. 4. Talbot images created by observing moiré interference at ¼, ½, ¾, and 1 Talbot distance. Incident angle is α.

In this work, Talbot images are investigated when ultrasound waves are applied vertically, or at α=0˚. Pressure Acoustics, Frequency Domain interface in COMSOL is used to model the stationary acoustic field in the water, to obtain the acoustic intensity distribution. The acoustic wave propagation is assumed to be linear, since the amplitude of shear waves in the propagation domain is much smaller than that of the pressure waves. Therefore, it is safe to neglect the shear waves and the nonlinear effects. By identifying the acoustic pressure field, the acoustic intensity field is defined.

The objective is to investigate the formation of Talbot images using ultrasound, employing Comsol. The set up used is shown in Fig. 5. In this figure, a sound source vertically applies ultrasound wave on a grating, where α=0˚. The Talbot images at $1D_T^\alpha$, $\frac{3}{4}D_T^\alpha$, $\frac{1}{2}D_T^\alpha$, and $\frac{1}{4}D_T^\alpha$ distance are obtained, Fig.6. The parameters used in the simulation is shown in Table I. Using an ultrasound frequency of 3.5 MHz results is a wave length of 0.42 mm with a wave speed of 1483 in water. This means that the wavelength is less than 10% of the pitch size of 4.7 mm. In order to ensure acceptable accuracy, wave length should be much smaller than the size of the pitch.

Fig. 6 shows the interference image at any distance from 0 to 1Talbot distance. This image is unique, since it cannot be obtained experimentally, since the experimental images are cross sections images at specific distances. Fig. 6 shows how the contours evolve with distance, as the detector moves far from the grating along the wave propagation. From the figure, it can be realized that the image at $1D_T^\alpha$ is the exact replica of the grating, which is placed at 0 distance. This is realized by comparing the peaks at $1D_T^\alpha$ and at 0. However, at $\frac{1}{2}D_T^\alpha$, the image is shifted, parallel to the grating, by ½ period. This image is called complementary image, and this is in agreement with Talbot theory [1, 6, 7]. The simulation confirms that moiré interference takes place in acoustic waves. The figure shows also that the frequency of the peaks doubles at $\frac{1}{4}D_T^\alpha$ and $\frac{3}{4}D_T^\alpha$ distances. This is also in agreement with Talbot theory. Using the simulation results of this figure, one can obtain the pattern at any distance. It should be also noted that at the sides of the tanks, there is some distortion. This distortion is due to the discontinuity of the medium (water).

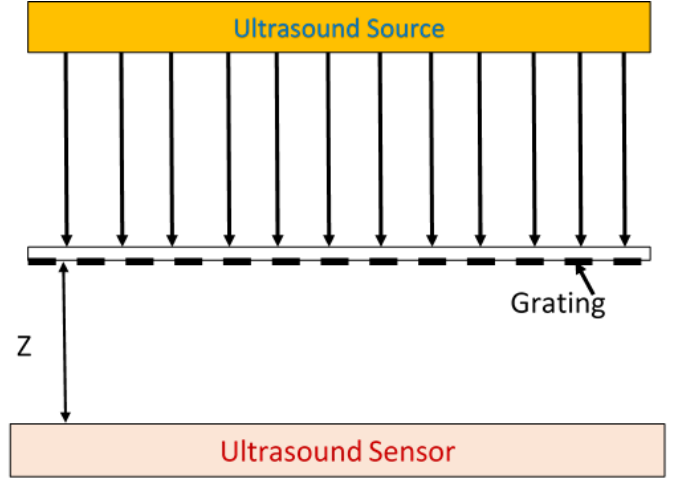

Fig. 5. Set-up used simulation for capturing the grating image at different Talbot distance.

TABLE I.
PARAMETERS USED IN SIMULATION

| Parameter (unit) | Value |
|---|---|
| Source frequency $f_0$ (MHz) | 3.5 |
| speed of wave (m/s) | 1483 |
| Wavelength λ (mm) | 0.42 |
| Pitch size of diffraction grating g (mm) | 4.7 |
| Height of water container (mm) | 72 |
| Width of water container (mm) | 120 |
| Talbot distance Dt (mm) | 102.5 |

The sound pressures at $1D_T^\alpha$ along the x-axis of Fig. 6 is plotted in Fig. 7. Although there are no sharp well defined peaks (maxima), there is a general profile for peaks. Every maximum in this figure represents a moiré fringe. As can be seen, the peaks are roughly separated by about 4-5 mm. The lack of peak sharpness can be attributed to the paraxial and Fresnel approximation, since λ/g is not very small and N is not very large. Fig. 8 shows the simulation results of the intensity profile along the x axis of Fig. 6, at 1 Talbot distance. Each peak represents a Talbot fringe. Again, the lack of sharpness is attributed to the diffraction because of paraxial and Fresnel approximation and noise. This figure shows more clearly that the separation is about 4.7 mm.

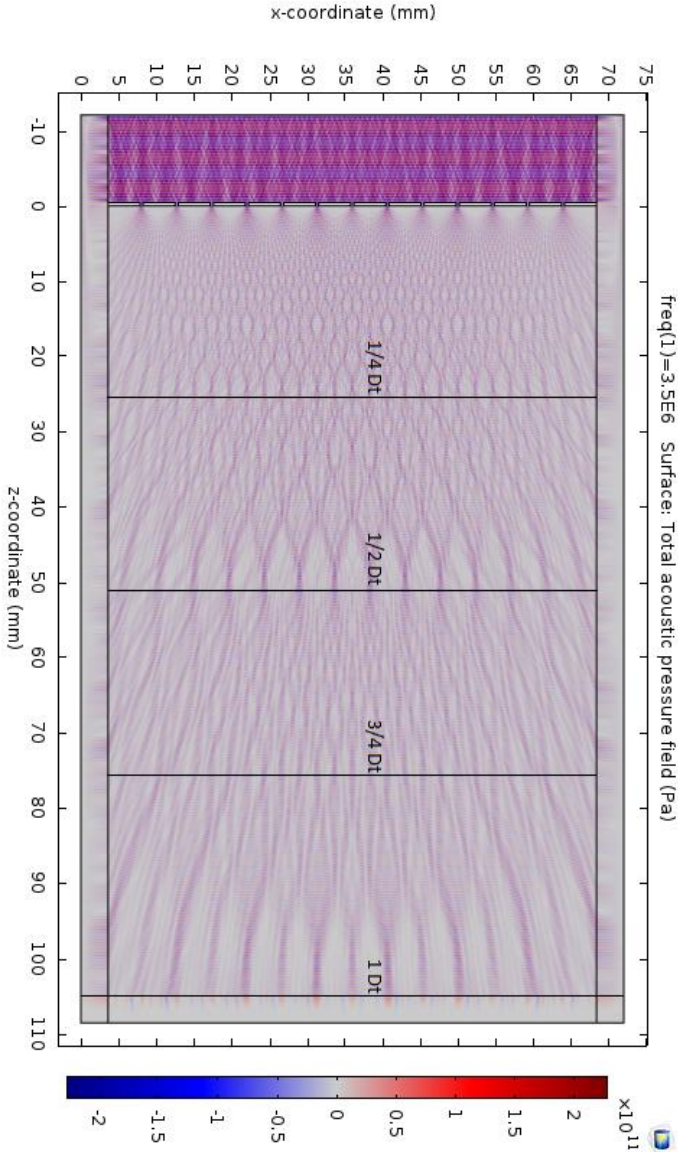

Fig. 6. Diffraction pattern showing interference images from 0 to $1D_T^\alpha$.

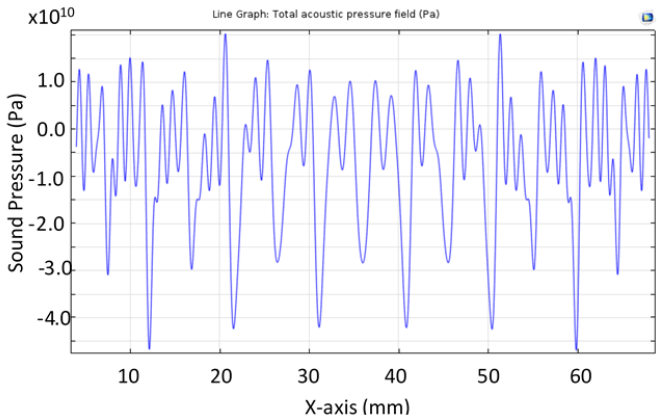

Fig. 7. Simulated sound pressure plot along the x-axis at $1D_T^\alpha$.

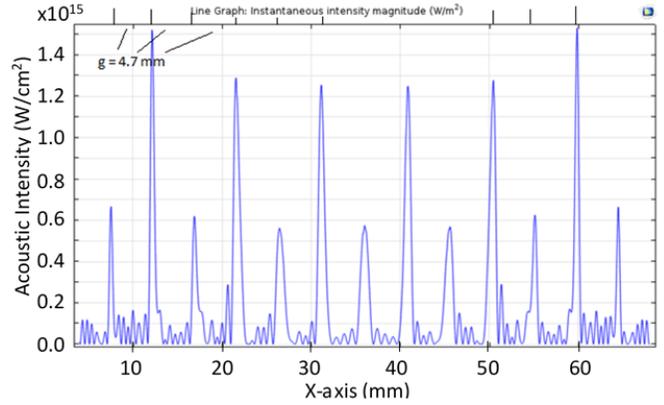

Fig. 8. Simulated ultrasound intensity along the x-axis at $1D_T^\alpha$. Distance between peaks is about 4.7 mm which is equal to g.

## IV. CONCLUSIONS

In this study, Comsol multi-physics program is used to investigate acoustic moiré interference. The results confirmed the experimental results and proved conclusively that moiré interference exists in acoustic waves as it does in optical waves. Comsol study shows a picture of interference along the propagation of the wave, Fig. 6. This image cannot be easily obtained experimentally, without taking huge number of pictures to reconstruct the full image, as CT scan does. Because acoustic waves can penetrate many materials, it can be used to image the interior of the object, including human body. Future work should concentrate on the development and testing of acoustic moiré interference systems. This system can provide many advantages over optical systems, including, but not limited to, safety and penetration ability. Acoustic moiré can be used in medical imaging, surface inspection, bulk material inspection, and other applications.